\documentclass{raa}
\usepackage{graphicx,times,rotating}
\usepackage{graphicx,subfigure}
\usepackage[]{natbib}
\newcommand\exo{EXO~2030+375}
\usepackage{amssymb,amsmath}
\bibpunct{(}{)}{;}{a}{}{,}
\input{epsf.sty}
\input{psfig.sty}
\usepackage{hyperref}

\hypersetup{pdftitle = The title of my PDF, pdfauthor = My name, pdfsubject= The subject, pdfkeywords = keyword1 keyword2 keyword3}
\hypersetup{colorlinks = true, linkcolor = green, anchorcolor = red, citecolor = blue, filecolor = red, pagecolor = red, urlcolor = red}

\begin{document}

\title{Suzaku Observation of Be/X-ray Binary Pulsar EXO~2030+375}
 \volnopage{ {\bf 2014} Vol.\ {\bf X} No. {\bf XX}, 000--000}
   \setcounter{page}{1}

\author{Sachindra Naik \and Gaurava K. Jaisawal}

\institute{Astronomy and Astrophysics Division, Physical Research Laboratory, Ahmedabad, India ;
              {\it snaik@prl.res.in}\\
	  \vs \no
	   {\small Received ...; accepted ...}
}

\abstract{In this paper we study the timing and spectral properties of Be/X-ray binary pulsar 
EXO~2030+375 using a $Suzaku$ observation on 2012 May 23, during a less intense Type~I 
outburst. Pulsations were clearly detected in the X-ray light curves at a barycentric 
period of 41.2852 s which suggests that the pulsar is spinning-up. The pulse profiles 
were found to be peculiar e.g. unlike that obtained from the earlier $Suzaku$ observation 
on 2007 May 14. A single-peaked narrow profile at soft X-rays (0.5-10 keV range) changed 
to a double-peaked broad profile in 12-55 keV energy range and again reverted back to a 
smooth single-peaked profile at hard X-rays (55-70 keV range). The 1.0-100.0 keV broad-band 
spectrum of the pulsar was found to be well described by three continuum models such as (i) 
a partial covering high energy cut-off power-law model, (ii) a partially absorbed power-law 
with high-energy exponential rolloff and (iii) a partial covering Negative and Positive power 
law with EXponential (NPEX) continuum model. Unlike earlier $Suzaku$ observation during which 
several low energy emission lines were detected, a weak and narrow Iron K$_\alpha$ emission 
line at 6.4 keV was only present in the pulsar spectrum during the 2012 May outburst. Non-detection
of any absorption like feature in 1-100 keV energy range supports the claim of absence of cyclotron
resonance scattering feature in EXO~2030+375 from earlier $Suzaku$ observation. Pulse-phase resolved 
spectroscopy revealed the presence of additional dense matter causing the absence of second 
peak from the soft X-ray pulse profiles. The details of the results are described in the paper.
}

\authorrunning{S. Naik \& G. K. Jaisawal}
\titlerunning{Suzaku Observations of EXO~2030+375}

\maketitle

\section{Introduction}

X-ray Binaries are known to be strong X-ray emitters and appear as the brightest 
X-ray sources in the sky. Depending on the mass of the optical companion, X-ray 
binaries are classified as low mass X-ray binaries (LMXBs) and high mass X-ray 
binaries (HMXBs). Based on the type of optical companion, the HMXBs are further 
classified as Be/X-ray binaries (largest subclass of HMXBs) and supergiant X-ray 
binaries. Though evolutionary model calculations show that binary systems with 
white dwarf and Be star or black hole and Be star should also exist, observational 
evidence on the existence of such binary systems are faraway (Zhang, Li \& Wang 2004 
and references therein). However, recently discovered Be/X-ray binary system with a 
black hole as the X-ray source (MWC~656; Casares et al. 2014) corroborates the model 
calculations. X-ray emitting compact object in most of the Be/X-ray binaries is generally 
a neutron star where as the optical companion is a B or O-type star that shows Balmer 
emission lines in its spectra. The neutron star in these binary systems is typically in a 
wide orbit with moderate eccentricity. The orbital period of these systems is in the range 
of 16-400 days. 

X-ray emission in the Be/X-ray binary systems is known to be due to the accretion of 
mass from the Be circumstellar disk on to the neutron star at the periastron passage. 
The abrupt accretion of huge amount of mass on to the neutron star results in strong 
X-ray outbursts (Okazaki \& Negueruela 2001) enhancing the source luminosity by a factor 
of more than $\sim$10 or more. Pulsars in these systems show periodic normal (Type I) 
X-ray outbursts that coincide with the periastron passage of the neutron star and giant 
(Type II) X-ray outbursts which do not show any clear orbital dependence (Negueruela et 
al. 1998). The spin period of these pulsars is found to be in the range of a few seconds 
to several hundred seconds. The X-ray spectra of these pulsars are generally hard. 
Fluorescent iron emission line at 6.4 keV is observed in the spectrum of most of the 
accretion powered X-ray pulsars. Cyclotron resonance scattering features (CRSF) have 
been detected in the broad-band X-ray spectrum of some of these pulsars. Detection of 
CRSF provides direct estimation of surface magnetic field of these objects. For a 
brief review of the properties of the transient Be/X-ray binary pulsars, refer to 
Paul \& Naik (2011).

Transient Be/X-ray binary pulsar \exo~ was discovered during a giant outburst 
in 1985 with $EXOSAT$ observatory (Parmar et al. 1989a). Optical and near-infrared 
observations identified a B0 Ve star as the counterpart of \exo~ (Motch \& Janot-Pacheco 
1987; Coe et al. 1988). Using the {\it EXOSAT} observation during the giant outburst, 
spin and orbital period of the binary pulsar were estimated to be 42 s and 44.3-48.6 
days, respectively (Parmar et al. 1989a). During the declining phase of the giant 
outburst, the pulsar was found to be dramatically spinning-up at a timescale of
$-P/\dot{P} \sim$ 30 yr. Strong luminosity dependence of the pulse profile of 
\exo~ was detected during the giant outburst in 1985 (Parmar et al. 1989b). The 
pulse profile was characterized by the presence of two peaks which were separated 
by $\sim$180$^\circ$ phase. The strength of the peaks in the pulse profile reversed 
as the pulsar luminosity decreased by a factor of $\sim$100. This was explained in 
terms of change in the pulsar emission from a fan-beam to a pencil-beam as the 
luminosity decreased resulting in the interchange in the strength of the main-pulse 
and inter-pulse. {\it Suzaku} observation of the pulsar at the peak of a Type~I
outburst, however, showed that the shape of pulse profiles was complex due to
the presence of prominent dips at several pulse phases (Naik et al. 2013). The 
dips were found to be strongly energy dependent and were present up to as high 
as $\sim$70 keV. An extensive monitoring of \exo~ with $BATSE$ and Rossi X-ray 
Timing Explorer ($RXTE$) showed that a normal outburst has been detected for 
nearly every periastron passage for $\sim$13.5 years (Wilson, Fabregat \& Coburn 
2005). Using {\it BATSE} observation of a series of consecutive Type~I outbursts 
of the pulsar, Stollberg et al. (1997) derived the orbital parameters of the binary 
system.

The spectrum of the pulsar, obtained from {\it EXOSAT} observation during
1985 giant outburst was described by a composite model consisting of a blackbody
component with temperature $\sim$1.1 keV and a power law component describing
the hard X-ray part (Reynolds, Parmar \& White 1993; Sun et al. 1994). However, 
a model consisting of a blackbody and a power-law with an exponential cut-off was 
required to fit the 2.7--30 keV spectrum obtained from the {\it RXTE} observation 
of the pulsar during the 1996 June-July outburst (Reig \& Coe 1999). Possible 
detection of cyclotron resonance scattering features at $\sim$36 keV (Reig \& 
Coe 1999), $\sim$11 keV (Wilson et al. 2008), $\sim$63 keV (Klochkov et al. 2008) 
in the spectrum of the pulsar have been reported earlier. However, the absence of 
any such feature in 1-100 keV spectrum of the pulsar ruled out the earlier suggestions 
of the detection of the cyclotron line in \exo~ (Naik et al. 2013). The broad-band
{\it Suzaku} spectrum of the pulsar was best fitted with a partial covering high-energy 
cutoff power-law model. As the pulsar was very bright, several low-energy emission
lines were also detected in the spectrum (Naik et al. 2013). 

{\it Suzaku} observation of the pulsar at the peak of a Type~I outburst of a 
significantly reduced intensity was used for a detailed study of the evolution 
and luminosity dependence of the absorption dips and corresponding changes in
its spectral features. The results obtained from this study are described in
this paper.

\section{Observation}
\begin{figure*}
\centering
\includegraphics[height=12.0cm,width=7.0cm,angle=-90]{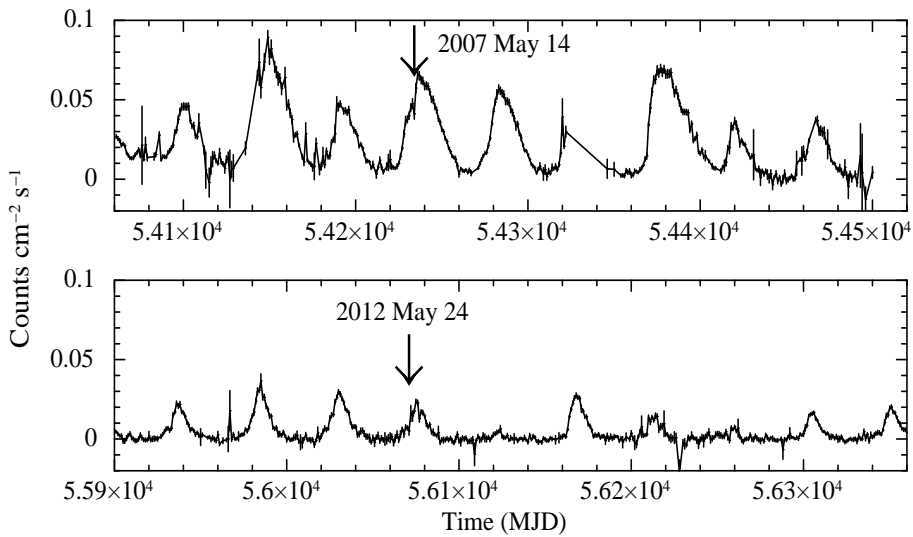}
\caption{$Swift$/BAT light curves of EXO~2030+375 in the 15-50 keV energy band, 
from 2006 November 21 (MJD 54060) to 2008 February 04 (MJD 54500) and 2011 
December 05 (MJD 55900) to 2013 March 09 (MJD 56360) in top and bottom panels,
respectively. The arrow marks in both the panels show the date of the Suzaku 
observations of the pulsar during its Type~I outbursts.} 
\label{f1}
\end{figure*}

The transient Be/X-ray binary pulsar \exo~ shows bright Type~I X-ray outbursts 
consistently at each periastron passage of the neutron star. The luminosity
of the pulsar at the peak of these Type~I outbursts are, however, found to be 
different. \exo~ was observed on 2007 May 14 and 2012 May 23 with the
instruments onboard {\it Suzaku} when the pulsar was undergoing X-ray outbursts.
During both the {\it Suzaku} observations, the peak luminosity was significantly
different. One-day averaged light curve of \exo~ in 15-50 keV energy range obtained
from the {\it Swift}/BAT monitoring data from 2006 November 21 to 2008 February 04 
(top panel) and 2011 December 05 to 2013 March 09 (bottom panel) covering both the
Type~I outbursts are shown in Figure~\ref{f1}. The arrow marks in both the panels
show the {\it Suzaku} observations of the pulsar. The results obtained from the
2012 May {\it Suzaku} observation of the pulsar during a significantly 
less intense Type~I outburst are discussed in this paper. The observation was
carried out at the ``XIS nominal'' pointing position for a total exposures of 
$\sim$78 ks and $\sim$72.5 ks for the X-ray Imaging Spectrometer (XIS) and Hard 
X-ray Detector (HXD), respectively. The XIS was operated in the ``1/4 window'' 
option covering a 17$'$.8$\times$4$'$.4 field of view. 

{\it Suzaku}, the fifth Japanese X-ray astronomy satellite, was launched on 2005
July 10 (Mitsuda et al. 2007). It has two sets of detectors such as X-ray Imaging 
Spectrometer (XIS : Koyama et al. 2007) and Hard X-ray Detectors (HXD : 
Takahashi et al. 2007). Installed at the focal plane of four X-ray telescopes 
(XRT : Serlemitsos et al. 2007), the XIS consists of three front-illuminated CCD 
cameras (XIS-0, 2 \& 3) and one back-illuminated CCD camera (XIS-1) sensitive in 
0.4-12 keV and 0.2-12 keV energy ranges, respectively. The non-imaging instrument 
HXD acquires data in 10-70 keV range with the Si~PIN photo diodes and in 40-600 keV 
range with the GSO scintillators. Combining XIS with HXD, {\it Suzaku} covers a broad 
energy band for the study of X-ray sources. The publicly available archival data 
(version 2.7.16.33) of the 2012 May observation of \exo~ were used in the present 
work. As XIS-2 was nonoperational during above observation of the pulsar, data 
from other three XIS, PIN and GSO were used in our analysis.

\section{Analysis and Results}

We used HEASoft software package (version 6.12) in our analysis. The calibration
database files released in 2012 February 10 (for XIS) and 2011 September 13 (for
HXD) by the instrument teams are used for data reduction. The unfiltered XIS and 
HXD event data were reprocessed by using the $``aepipeline''$ package of HEASoft.
Barycentric correction was applied to the reprocessed XIS and HXD event data by
using the $``aebarycen''$ task of FTOOLS. These barycentric corrected XIS and HXD 
event files were used for further analysis. Source light curves and spectra were 
accumulated from the reprocessed XIS cleaned event data by selecting circular 
region with a 3$'$ diameter around the central X-ray source. The background
light curves and spectra were extracted from these event data by selecting circular
regions away from the source position. {\it ``xisrmfgen''} and {\it ``xissimarfgen''}
tasks were used to generate response files and effective area files for corresponding
XIS detectors. Hard X-ray light curves and spectra of the pulsar were extracted
from the reprocessed HXD/PIN and HXD/GSO event data by using the {\it ``XSELECT''}
task of FTOOLS. Simulated background event data for HXD/PIN and HXD/GSO, provided 
by the instrument team were used to estimate background light curves and spectra 
for {\it Suzaku} observation of the pulsar. Response files released in 
2011 June (for HXD/PIN) and 2010 May (for HXD/GSO) and an effective area file 
released on 2010 May for HXD/GSO were used for spectral analysis.

\subsection{Timing Analysis}
Source light curves with time resolutions of 2 s, 1 s and 1 s were extracted
from the barycenter corrected XIS-0 (in 0.4-12 keV energy range), PIN (in 10-70 
keV energy range) and GSO (in 40-200 keV energy range) event data, respectively.
As described earlier, the background light curves were extracted for XIS, PIN
and GSO detectors and subtracted from the source light curves. By applying pulse 
folding and $\chi^2$ maximization technique, the spin period of the pulsar was 
estimated to be 41.2852(3) s. The estimated value of the pulse period of
the pulsar showed a global spin-up trend while comparing with the earlier reported
values. Earlier reported values of spin period of the pulsar at different epochs are 
tabulated in Table~\ref{sp}. Though spin-up, spin-down and constant spin frequency 
episodes are observed at smaller time scales (Parmar et al. 1989a; Wilson et al. 2005; 
2008), the pulsar showed overall spin-up trend at longer time scale.

\begin{table} 
\scriptsize
\caption{Global spin period history of EXO~2030+375}
\label{table}
\centering
\begin{tabular}{|l | c | c | c |}
\hline
Date of Observation			& MJD         		& Spin period & References \\
\hline
1985 May 18 - 1985 Nov 03$^a$   	&46203 - 46372		&41.8327 s - 41.7275 s  &Parmar et al. (1989a) \\
1991-2003$^b$          			&48400 - 52900 		&41.6910 s - 41.6736 s  &Wilson et al. (2005) \\
2006 June 22 - 2006 Nov 11$^c$  	&53908 - 54050		&41.6320 s - 41.4421 s  &Wilson et al. (2008) \\
2007 May 14             &54234		&41.4106 s              &Naik et al. (2013) \\
2012 May 24             &56071		&41.2852 s              &present study \\
\hline
\end{tabular}
\begin{flushleft}
$^a$ : Maximum and minimum values of spin period out of 13 measurements (Table 1 of Parmar et al. 1989a), $^b$ : Maximum and minimum values of spin period out of 55 measurements (Figure 1 of Wilson et al. 2005),
$^c$ : Maximum and minimum values of spin period out of 29 measurements (Figure 4, top panel of Wilson et al. 2008).
\end{flushleft}
\label{sp}
\end{table}

\begin{figure}
\centering
\includegraphics[height=7.0cm,width=10.0cm,angle=-90]{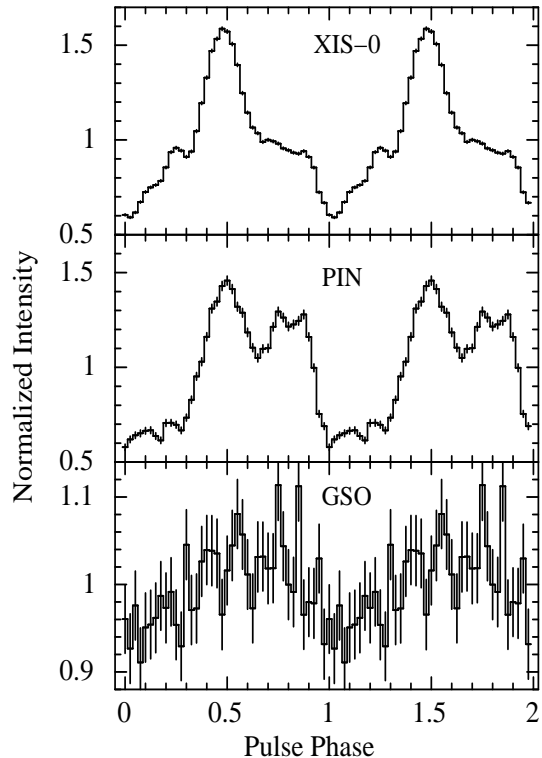}
\caption{Pulse profiles of EXO~2030+375 in 0.4-12 keV range (XIS-0; top panel), 
10-70 keV range (PIN; middle panel) and 40-200 keV range (GSO; bottom panel), 
obtained from the background subtracted light curves by using the estimated 
41.2852 s pulse period. The errors in the figure are estimated for the 1$\sigma$ 
confidence level. Two pulses are shown for clarity.}
\label{f2}
\end{figure}

The pulse profiles of the pulsar were obtained by 
folding the background subtracted light curves obtained from  XIS, PIN and GSO 
data with the estimated spin period and are shown in Figure~\ref{f2}. It can be
seen from this figure that the pulse profiles are significantly different at
different energy ranges. The shape of the pulse profiles obtained from the
2012 May {\it Suzaku} observation of the pulsar was found to be significantly
different to that reported from earlier observations (Naik et al. 2013 and
references therein). A single-peaked profile in 0.4-12 keV range (top panel)
changed to a structured double-peaked profile in 10-70 keV range (middle panel) 
which again became smooth and single-peaked in 40-200 keV range (bottom panel).

\begin{figure*}
\centering
\includegraphics[height=14.cm,width=11.5cm,angle=-90]{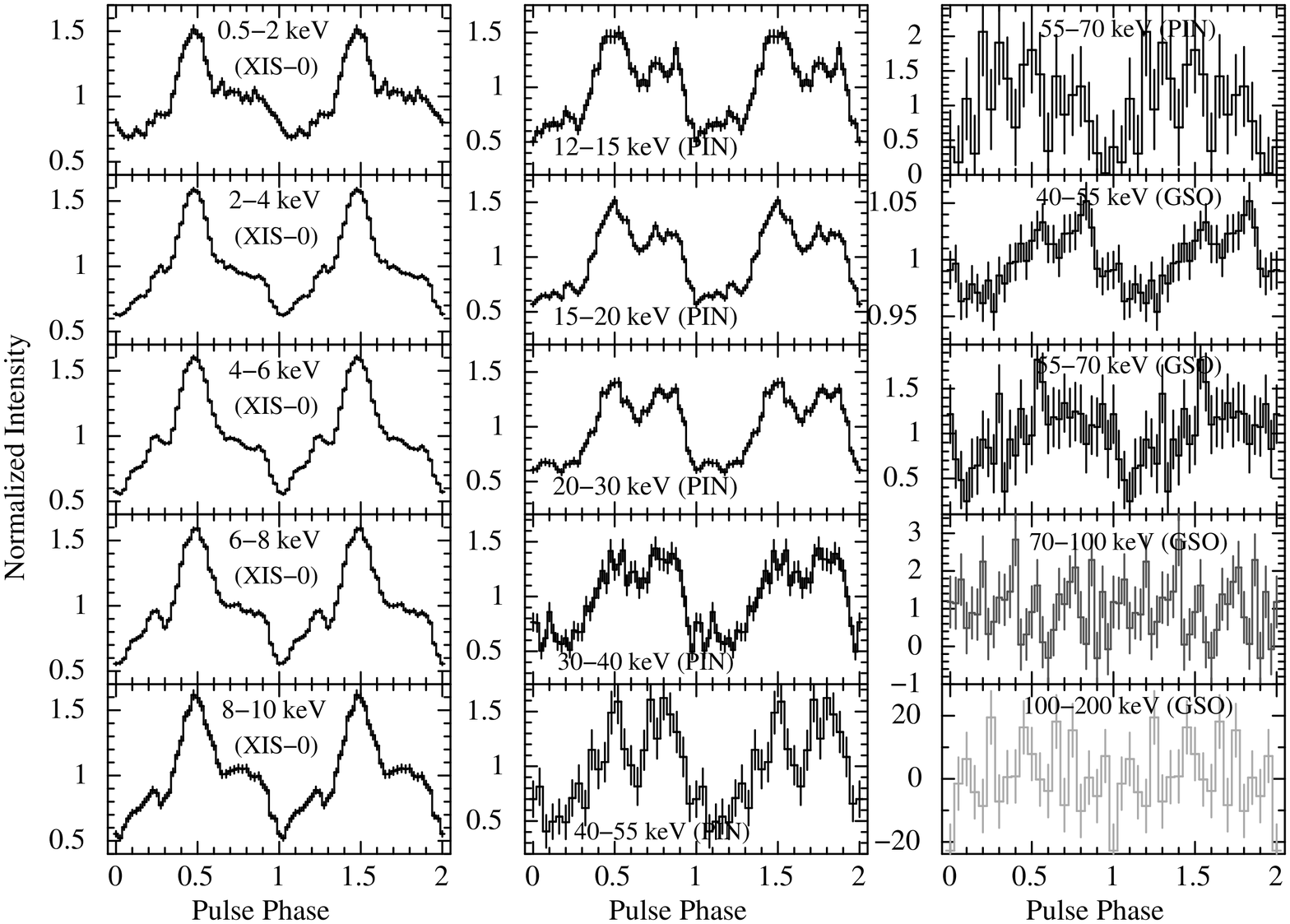}
\caption{Energy-resolved pulse profiles of EXO~2030+375 obtained from XIS-0, 
HXD/PIN and HXD/GSO light curves at various energy ranges. The presence of
absorption dips in profiles at higher energies can be seen in 0.6--0.8 
pulse phase range. The error bars represent 1$\sigma$ uncertainties. Two 
pulses in each panel are shown for clarity.}
\label{f3}
\end{figure*}

To investigate the evolution of the pulse profile with energy, light curves
at various energy ranges were extracted from the XIS, PIN and GSO event data.
Background light curves in same energy ranges were also extracted and subtracted
from the corresponding source light curves. Energy resolved pulse profiles were
generated by using the estimated spin period of the pulsar and shown in 
Figure~\ref{f3}. It can be seen that the pulse profile of the pulsar was 
single-peaked up to $\sim$8 keV beyond which a hump-like structure appeared
after the main peak. The pulse profile became double-peaked up to $\sim$40 keV
beyond which it again became single-peaked. The 41.2852 s pulsation was seen
in HXD/PIN and HXD/GSO light curves up to $\sim$70 keV range, beyond which 
it was absent. Though energy dependent pulse profiles are seen in EXO~2030+375,
the profiles obtained during the 2012 May outburst are significantly different
to that obtained during 2007 May outburst. Though the observations were carried
out during Type~I outbursts with the detectors onboard {\it Suzaku}, entirely 
different type of pulse profiles require a detailed spectral investigation to 
understand the emission mechanism in the pulsar.

\subsection{Spectral Analysis}
\subsubsection{Pulse-phase-averaged spectroscopy}

We carried out simultaneous spectral fitting of the XIS (XIS-0, XIS-1 
and XIS-3), PIN and GSO data to investigate the energy dependence of 
pulse profile seen in 2012 May {\it Suzaku} observation of EXO~2030+375. 
As described earlier, source spectra were extracted from the XIS, PIN 
and GSO event data and corresponding background spectra and response files 
were obtained by following appropriate procedure. Using the extracted source
spectra, background spectra and response files, simultaneous spectral 
fitting was carried out using the software package XSPEC v12.7.1.
Because of the presence of known structures in the XIS spectra at Si and Au
edge, data in the 1.7-1.9 keV and 2.2-2.4 keV range were ignored from the
spectral fitting. XIS spectra were re-binned by a factor of 6 from 1 to 10 
keV whereas the HXD/PIN spectrum was re-binned by a factor of 4 from 23 keV
to 45 keV and by a factor of 6 from 45 keV to 70 keV. The binning of GSO 
spectrum, however, was done as suggested by the instrument 
team\footnote{http://heasarc.gsfc.nasa.gov/docs/suzaku/analysis/gsobgd64bins.dat}. 
In the spectral fitting, all the model parameters were tied together except for 
the relative instrument normalizations. The broad-band energy spectra in 1-100 keV 
energy range were fitted with several continuum models such as (i) high-energy cutoff 
power-law model, (ii) a power-law with high energy exponential rolloff, and 
(iii) Negative and Positive power law with EXponential cutoff (NPEX) continuum 
model. The analytical form of above continuum models are

High energy cutoff power law model - \\
\begin{equation}
I(E)= \left\{ \begin{array}{lc}
E^{-\gamma} & (E\le E_c) \\
E^{-\gamma}\exp\left(-\frac{E-E_c}{E_f}\right) & (E>E_c)
\end{array}\right.
\end{equation}
where $E_c$ and $E_f$ are cutoff energy and  folding energy, respectively.
\\
\\
High energy exponential rolloff model - \\
\begin{equation}
\centering
I(E) = KE^{-\alpha} e^{-E/kT}
\end{equation}

where, $K$ is normalization constant and $\alpha$ is photon index. $kT$ represents 
cutoff-energy of power-law in unit of keV.
\\
\\
NPEX continuum model  -\\
\begin{equation}
\centering
NPEX(E) = ( N_1 E^{-\alpha_1} + N_2 E^{+\alpha_2} ) ~ e^{-E/kT}
\label{e1}
\end{equation}
where $E$ is energy of X-ray photons, $N_1$, $N_2$, $\alpha_1$, $\alpha_2$ are normalization and index of negative and positive power-law, respectively. $kT$ represents cutoff-energy of power-law in unit of keV.

\begin{figure}
\centering
\includegraphics[height=9.0cm,width=7.0cm,angle=-90]{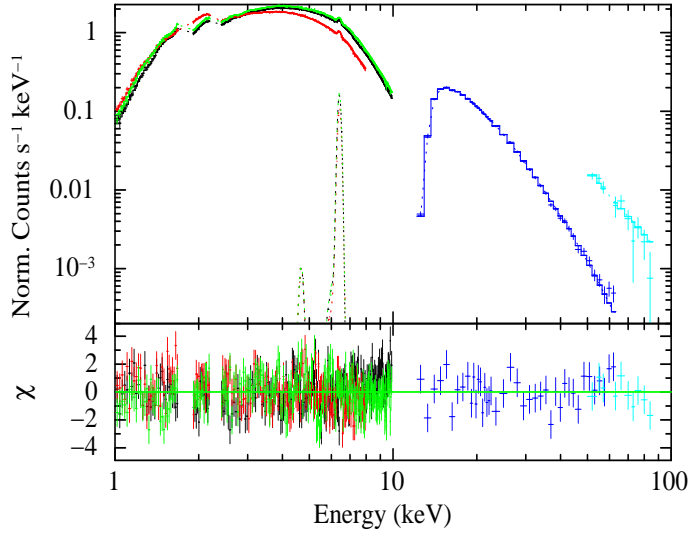}
\caption{Energy spectrum of EXO~2030+375 obtained with the XIS-0, XIS-1, XIS-3, 
PIN, and GSO detectors of the {\it Suzaku} observation during the 2012
May-June Type~I outburst, along with the best-fit model comprising a partially 
absorbed power law with a high-energy cutoff power law continuum model, a Gaussian function
for the narrow iron emission line along with the interstellar absorption. The
contributions of the residuals to the $\chi^2$ for each energy bin for the 
best-fit model are shown in the bottom panel.}
\label{spec_highecut}
\end{figure}

\begin{figure}
\centering
\includegraphics[height=9.0cm,width=7.0cm,angle=-90]{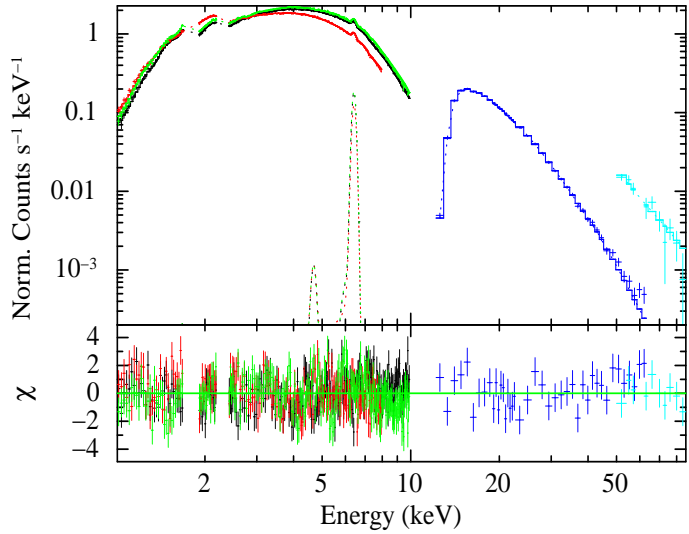}
\caption{Energy spectrum of EXO~2030+375 obtained with the XIS-0, XIS-1, XIS-3, 
PIN, and GSO detectors of the {\it Suzaku} observation during the 2012
May-June Type~I outburst, along with the best-fit model comprising a partially 
absorbed power law with high-energy exponential rolloff model, a Gaussian function
for the narrow iron emission line along with the interstellar absorption. The
contributions of the residuals to the $\chi^2$ for each energy bin for the 
best-fit model are shown in the bottom panel.}
\label{spec_cutoffpl}
\end{figure}

\begin{figure}
\centering
\includegraphics[height=9.0cm,width=7.0cm,angle=-90]{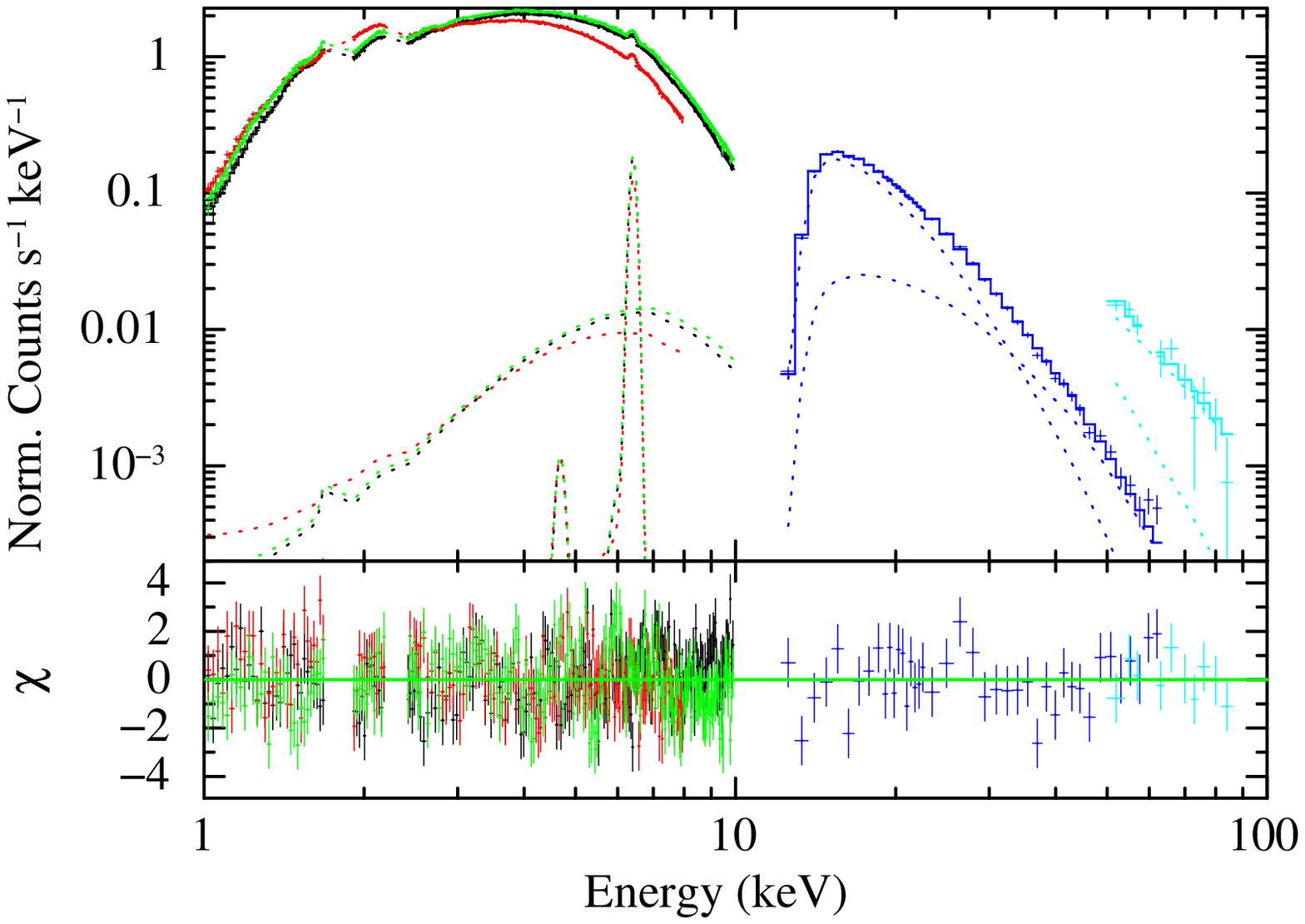}
\caption{Energy spectrum of EXO~2030+375 obtained with the XIS-0, XIS-1,
XIS-3, PIN and GSO detectors of the {\it Suzaku} observation during the 
2012 May-June Type~I X-ray outburst. The data are plotted with the
best-fit model comprising a partially absorbed NPEX continuum model and a 
Gaussian function for the narrow iron emission line along with the 
interstellar absorption. The contributions of the residuals to the 
$\chi^2$ for each energy bin for the best-fit model are shown in the 
bottom panel.}
\label{spec_npex}
\end{figure}

Additional components such as photoelectric absorption, a Gaussian function 
for an iron emission line were added to the continuum models while fitting
the pulsar spectra. It was found that all the continuum models provided similar
fits to the pulsar spectra with reduced $\chi^2$ of $\sim$1.7. As in case of 
other Be/X-ray binary pulsars, a partial covering absorption component {\it pcfabs} 
was then applied to above continuum models in the spectral fitting. Addition of
this component to above three continuum models improved the spectral fitting
significantly yielding a reduced $\chi^2$ of $\sim$1.4. The best-fit parameters 
obtained from the simultaneous spectral fitting to the XIS, PIN and GSO data are 
given in Table~\ref{spec_par}. The count rate spectra of the pulsar EXO~2030+375
are shown in Figure~\ref{spec_highecut} (for high-energy cutoff power-law model), 
Figure~\ref{spec_cutoffpl} (for power-law with high energy exponential rolloff model),
and Figure~\ref{spec_npex} (for NPEX continuum model) along with the model
components (top panels) and residuals to the fitted models (bottom panels). In the
spectral fitting using above models, there was no signature of presence of a cyclotron
resonance scattering feature (CRSF) at earlier reported energies in EXO~2030+375.

\begin{table*}
\centering
\caption{Best-fit parameters (with 1$\sigma$ errors) obtained from the spectral fitting
of the 1-100 keV spectra of the $Suzaku$ observation of EXO~2030+375 during the 2012
May Type~I outburst. Model-I, Model-II and Model-III represent the partially absorbed
power law with a high-energy cutoff continuum model with interstellar absorption and Gaussian components, a partially absorbed power law with high-energy exponential rolloff model with interstellar absorption and Gaussian components and a partially absorbed
NPEX continuum model with interstellar absorption and Gaussian components, respectively.}
\begin{tabular}{llllllllllll}
\hline
\hline
Parameter    &\multicolumn {3}{c}{Value}	 \\
             &Model-I     &Model-II        &Model-III  \\
\hline
N$_{H1}$ (10$^{22}$ atoms cm$^{-2}$)   &2.02$\pm$0.02    &2.02$\pm$0.02   &1.93$\pm$0.02\\
N$_{H2}$ (10$^{22}$ atoms cm$^{-2}$)   &4.47$\pm$0.15    &5.16$\pm$0.15   &4.70$\pm$0.17\\
Covering fraction                      &0.53$\pm$0.01    &0.50$\pm$0.01   &0.44$\pm$0.01\\
High energy cut-off (keV)              &6.52$\pm$0.13    &20.2$\pm$0.4    &10.6$\pm$0.6\\
E-fold energy (keV)                    &24.6$\pm$0.5     &--              &--\\
Power-law index                        &1.26$\pm$0.01    &1.07$\pm$0.01   &0.79$\pm$0.04\\
Iron line energy (keV)		       &6.41$\pm$0.01    &6.42$\pm$0.01   &6.42$\pm$0.01\\
Iron line width (keV)                  &0.01$\pm$0.01    &0.03$\pm$0.01   &0.03$\pm$0.01\\
Iron line equivalent width (eV)        &27$\pm$2         &33$\pm$2        &33$\pm$2 \\
1-10 keV flux$^a$                      &4.5$\pm$0.2      &4.6$\pm$0.1     &4.6$\pm$0.2\\
10-70 keV flux$^a$                     &10.6$\pm$0.3     &10.3$\pm$0.2    &10.6$\pm$1.4\\
Reduced $\chi^2$                       &1.36 (650 dof)   &1.41 (651 dof) &1.38 (650 dof)\\
Relative Inst. Normalization           &1.0/0.94/0.99/1.11/1.06    &1.0/0.94/0.99/1.04/1.11    &1.0/0.94/0.99/1.11/1.09\\
(XIS-0/XIS-1/XIS-3/PIN/GSO)\\
\hline
\end{tabular}
\begin{flushleft}
N$_{H1}$ = Equivalent hydrogen column density, N$_{H2}$ = additional hydrogen column density, $^a$ : in 10$^{-10}$  ergs cm$^{-2}$ s$^{-1}$ unit. Quoted source flux is not corrected for interstellar absorption. \\
\end{flushleft}
\label{spec_par}
\end{table*}

\subsubsection{Pulse-phase-resolved spectroscopy}

Accretion powered transient X-ray pulsars show complex pulse profiles in
soft X-ray bands that gradually become smooth and single-peaked at higher
energies. As the soft X-ray photons emitted from the polar caps of the pulsar
are generally most affected due to the absorption by matter distributed in 
the vicinity of the neutron star and in the interstellar medium, the shape
of the pulse profiles become complex which is not the case in higher energies.
However, the shape of the pulse profiles of EXO~2030+375 obtained from the 
2012 May {\it Suzaku} observation is rather smooth and single-peaked at soft
X-rays and complex in higher energies. Therefore, it is interesting to do
a detailed spectral study at narrow pulse phases of the transient pulsar 
during the 2012 May outburst. For the pulse phase resolved spectral study,
we used data from XIS (XIS-0, XIS-1 \& XIS-3) and HXD/PIN detectors. 
We did not include HXD/GSO data in phase-resolved spectroscopy because of 
the lack of sufficient number of photons at each phase bin of the pulsar. XIS 
and PIN spectra were accumulated into 20 pulse-phase bins by applying phase
filter in the FTOOLS task XSELECT. Background spectra, response matrices
used in the phase-averaged spectroscopy were also used in the phase-resolved
spectral analysis. As all three continuum models were yielding similar
results while fitting phase-averaged spectra, two of the three models 
(high-energy cutoff power-law and NPEX continuum models) were used for 
simultaneous spectral fitting to the phase-resolved spectra in 1-70 keV 
range. In the spectral fitting, the values of relative instrument normalizations
were fixed at the values obtained from the phase-averaged spectroscopy (as
given in Table~\ref{spec_par}). It was found that certain parameters such
as absorption column density (N$_{H1}$), iron line energy and line width 
did not show any significant variation over pulse phases of the pulsar. 
Therefore, these parameters were also fixed at the phase-averaged values.

\begin{figure*}
\centering
\includegraphics[height=14.cm,width=11.5cm,angle=-90]{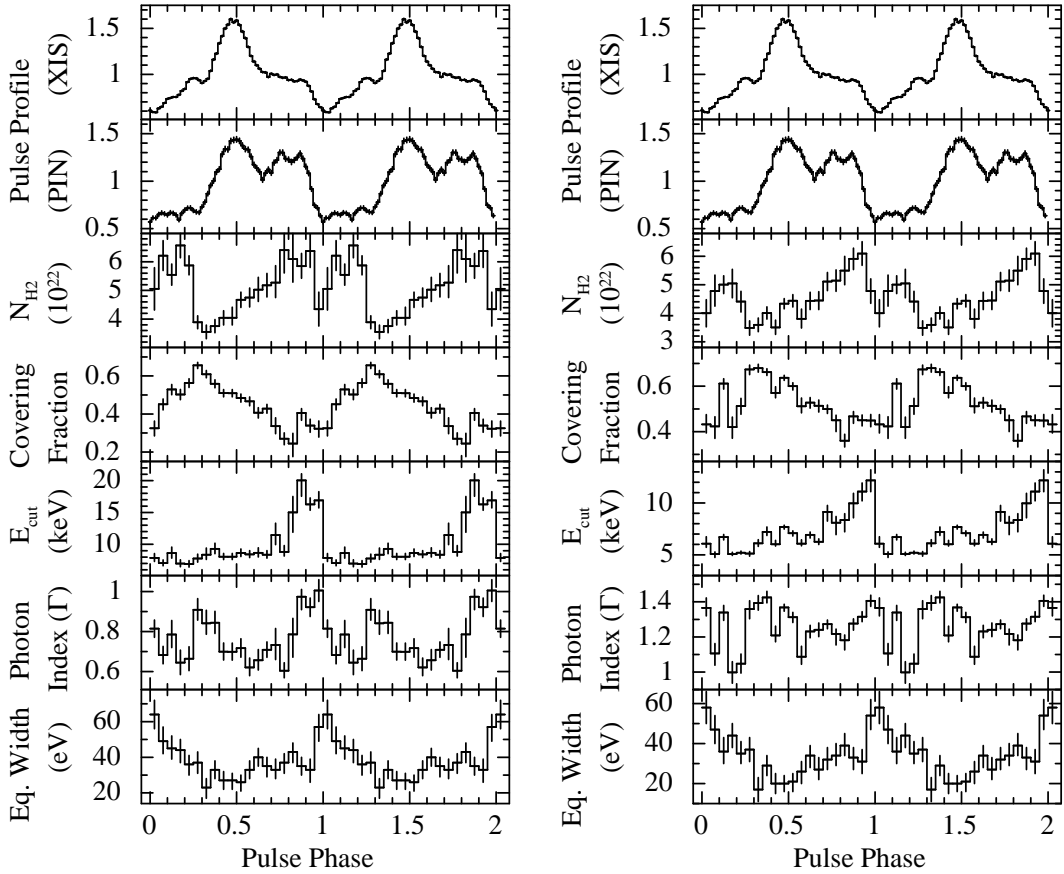}
\caption{Spectral parameters obtained from the pulse-phase-resolved spectroscopy
of {\it Suzaku} observation of EXO~2030+375. The XIS (in 0.4-12 keV range) and PIN
(in 10-70 keV range) pulse profiles are shown in top two panels of both sides of
the figure, respectively. The other panels show the spectral parameters obtained
by using a partial covering NPEX continuum model (left panels) and a partial covering
cutoff power law continuum model (right panels). The errors shown in the figure are
estimated for 1$\sigma$ confidence level.}
\label{ph_rs}
\end{figure*}

Parameters obtained from the simultaneous spectral fitting to the phase-resolved
XIS and PIN data in 1-70 keV range are shown in Figure~\ref{ph_rs}. Pulse profiles
obtained from XIS and PIN data are shown in top two panels on both the sides of
the figure. Parameters obtained from the spectral fitting using partially
absorbed NPEX and high-energy cutoff power-law continuum models along with
interstellar absorption and Gaussian function are shown in left and right 
panels of the figure, respectively. It can be seen that the parameters obtained 
from the phase-resolved spectral fitting using two different continuum models
followed similar pattern over pulse phases of the pulsar. In case of both 
the models, the value of additional column density (N$_{H2}$) was found to 
be high in 0.6-0.9 pulse phase range. High value of additional column density
can explain the absence of significant amount of soft X-ray photons in above 
pulse phase range. The absorption of soft X-ray photons by the additional
matter makes the pulse profile shallow in 0.6-0.9 pulse phase range (top panels
of Figure~\ref{ph_rs}). However, at hard X-rays, the effect of the additional
matter is drastically reduced making the pulse profile different compared to
that in soft X-ray bands. The pulsar spectrum was found to be marginally hard 
in 0.8-1.1 phase range along with high value of cutoff energy and iron line 
equivalent width. This coincides with the presence of a dip (primary dip in the
pulse profile) in the pulse profile at this phase range.

\section{Discussion}

The timing and spectral properties of transient Be/X-ray pulsar EXO~2030+375 have
been reported earlier (Naik et al. 2013 and references therein). During normal
Type~I outbursts, the pulse profile of the pulsar was found to be strongly energy
and luminosity dependent. At high luminosity, the pulse profile of the pulsar was
found to be complex because of the presence of several narrow prominent dips at
various spin phases (Naik et al. 2013). At low luminosity level, however, the 
pulse profile was smooth and single-peaked (Parmar et al. 1989). Strong luminosity
dependence of the pulse profile in EXO~2030+375 has been reported earlier by using
observations from several observatories (as discussed in Naik et al. 2013 and 
references therein). Though the pulsar was observed with {\it Suzaku} during two 
Type~I outbursts, the shape of the profiles obtained from these observations were
significantly different. During 2007 May Type~I outburst, the shape of the pulse 
profile was complex due to the presence of several energy dependent narrow dips at
various pulse phases. The strength of these dips gradually decreased with increase
in energy, making the hard X-ray profile smooth and single-peaked. However, during 
the 2012 May Type~I outburst, the shape of the profile was entirely different -- a
narrow single-peaked profile at soft X-rays which became a double-peaked profile up 
to $\sim$55 keV beyond which it again became single-peaked. Investigation of 
significant difference in the shape of pulse profiles during 2007 May and 2012 
May Type~I outbursts may provide information regarding the geometry of the 
matter distribution around the poles of the neutron star. Accretion of huge
amount of matter onto the neutron star (during bright X-ray outbursts in Be/X-ray
binary pulsars) causes changes in the geometry of the matter distribution around
the neutron star from a smooth accretion stream to several narrow accretion streams
that are phase-locked with the neutron star. These narrow streams of matter causes
several absorption dips in the pulse profiles during bright X-ray outbursts. However,
during low mass accretion duration, the observed pulse profiles are relatively 
smooth which is seen in case of 2012 May observation of EXO~2030+375 compared
to that during 2007 May outburst.

During 2007 May {\it Suzaku} observation, the pulsar was bright in X-ray compared to
that during 2012 May observation. The source flux in 1-70 keV range was estimated to be 
$\sim$8.9 $\times$ 10$^{-9}$ erg cm$^{-2}$ s$^{-1}$ (Naik et al. 2013). However, during 
the 2012 May observation, the pulsar was much fainter with estimated 1-70 keV flux 
of $\sim$1.5 $\times$ 10$^{-9}$ erg cm$^{-2}$ s$^{-1}$ (present work). It was also
evident from Figure~1 that the 2007 May Type~I outburst was much brighter in 15-50 keV 
range compared to the 2012 May Type~I outburst. While comparing spectral properties
of the pulsar, it was found that partial covering absorption model fitted well to the
broad-band {\it Suzaku} spectra during both the observations. The value of absorption
column density (N$_{H1}$) obtained from the spectral fitting was found to be comparable
in both the cases. However, the value of additional absorption column density (N$_{H2}$)
was about an order of magnitude higher during 2007 May observation compared to that 
during 2012 May observation. As the pulsar was bright during 2007 May observation, 
several narrow emission lines were also detected in the spectrum. Significantly
high value of additional absorption column density and presence of several narrow and
prominent absorption dips in the pulse profiles during the high luminosity level of
the pulsar indicate the accretion of huge amount of mass from the circumstellar 
disk of Be companion star during 2007 May outburst. Pulse-phase resolved spectroscopy
of 2007 May {\it Suzaku} observation revealed the presence of narrow streams of matter 
causing the absorption dips in the pulse profile and are phase locked with the pulsar. 
Lower value of additional column density, absence of absorption dips in the pulse 
profile of the pulsar and about an order of magnitude less luminous during 2012 May 
{\it Suzaku} observation confirm that mass accretion from the circumstellar disk
of the Be star to the neutron star was significantly low compared to that during
2007 May observation. 

In case Be/X-ray binary pulsars, it is known that the observed regular and 
periodic X-ray outbursts during the periastron passage of the neutron star are 
due to the evacuation of matter from the circumstellar disk of the Be star. Regular 
monitoring of these Be/X-ray binary pulsars showed that the peak luminosity of the 
neutron star during these outbursts varies with time e.g. in case of EXO~2030+375, 
peak luminosity of the pulsar varies by an order of magnitude between the 2007 May 
and 2012 May X-ray outbursts. The change in the peak luminosity during the outbursts 
can be explained as due to the difference in the amount of mass evacuated from the 
Be circumstellar disk by the neutron star during the periastron passage which in turn 
depends on the evolution of the Be circumstellar disk. Therefore, we suggest that the 
size of the circumstellar disk of the Be star in EXO~2030+375 binary system was 
relatively small during the 2012 May outburst compared to that during 2007 May 
observation.

Presence of dips in the pulse profile are seen in many transient Be/X-ray binary 
pulsars such as A~0535+262 (Naik et al. 2008), GRO~J1008-57 (Naik et al. 2011),
1A~1118-61 (Maitra et al. 2012) etc. The evolution of pulse profiles of Be/X-ray
binary pulsars from a smooth single-peaked profile (during quiescence) to a complex 
shape because of the presence of several prominent absorption dips (during Type~I 
outbursts) are briefly described in Paul \& Naik (2011) and Naik (2013) and references 
therein. Presence of the absorption dips in the pulse profiles of these transient Be/X-ray 
binary pulsars is explained as due to the abrupt accretion of huge amount of matter 
that disrupts the accretion stream into several narrow streams of matter that are 
phase locked with the neutron star. The presence of dips in the pulse profiles of
Be/X-ray transient pulsars during Type~I outbursts can be compared with the effect
of abrupt mass accretion onto weakly magnetized stars. Three-dimensional
Magnetohydrodynamic (MHD) simulations of mass accretion from the companion to 
weakly magnetized stars such as weakly magnetized neutron stars (millisecond pulsar),
magnetized white dwarfs in some cataclysmic variables etc. showed that during unstable
regime of mass accretion, the accreted matter can penetrate into the magnetosphere
leading to stochastic light curves (complex pulse profiles) whereas in stable accretion
regime, matter gets accreted in the form of streams yielding almost periodic light curves
(Romanova et al. 2008). In case of Be/X-ray binary pulsars, significant amount of mass
is being accreted onto the neutron star at the periastron passage, leading to X-ray outbursts. 
Depending on the amount of mass evacuated from the circumstellar disk of
the Be star, the pulse profile of the Be/X-ray binary pulsar gets modified as seen in
case of EXO~2030+375 -- a complex profile because of the presence of several narrow
and prominent dips at high luminosity level and a relatively smooth profile at low
luminosity level. Therefore, the change in shape of the pulse profiles with luminosity
in Be/X-ray binary pulsars agrees with the theoretical prediction by Romanova et al.
(2008) though it was done for low magnetic objects.

\section*{acknowledgments}
We thank the anonymous referee for his/her useful suggestions that improved the
manuscript. The research work at Physical Research Laboratory is funded by the 
Department of Space, Government of India. The authors would like to thank all the 
members of the {\it Suzaku} for their contributions in the instrument preparation, 
spacecraft operation, software development and in-orbit instrumental calibration. 
This research has made use of data obtained through HEASARC Online Service, provided 
by the NASA/GSFC, in support of NASA High Energy Astrophysics Programs.

\bibliographystyle{raa}

\end{document}